\documentclass[aps,prb,twocolumn,a4]{revtex4-1} 
\pdfoutput=1
\usepackage{amsmath,amssymb,amsfonts,upgreek}
\usepackage{color}
\usepackage{graphicx}
\usepackage{setspace}
\usepackage{bm}
\usepackage{comment}
\usepackage{hyperref}

\renewcommand{\vec}[1]{\bm{#1}}

\begin{document}

\title{On the calculation of crystal field parameters using Wannier
  functions}

\author{Andrea Scaramucci}
\email{andrea.scaramucci@mat.ethz.ch}
\affiliation{Materials Theory, ETH Z\"urich, Wolfgang-Pauli-Strasse
  27, 8093 Z\"urich, Switzerland}
\author{Jens Ammann}
\affiliation{Materials Theory, ETH Z\"urich, Wolfgang-Pauli-Strasse
  27, 8093 Z\"urich, Switzerland}
\author{Nicola A. Spaldin}
\affiliation{Materials Theory, ETH Z\"urich, Wolfgang-Pauli-Strasse
  27, 8093 Z\"urich, Switzerland}
\author{Claude Ederer}
\email{claude.ederer@mat.ethz.ch}
\affiliation{Materials Theory, ETH Z\"urich, Wolfgang-Pauli-Strasse
  27, 8093 Z\"urich, Switzerland}

\date{\today}

\begin{abstract}
We discuss the calculation of crystal field splittings using Wannier
functions and show how the ligand field contributions can be separated
from the bare Coulomb contribution to the crystal field by
constructing sets of Wannier functions incorporating different levels
of hybridization. We demonstrate this method using SrVO$_3$ as a generic
example of a transition metal oxide. We then calculate trends in the
crystal field splitting for two series of hypothetical tetragonally
distorted perovskite oxides and discuss the relation between the
calculated ``electro-static'' contribution to the crystal field and
the simple point charge model. Finally, we apply our method to the
charge disproportionated 5$d$ electron system CsAuCl$_3$. We show that
the negative charge transfer energy in this material leads to a
reversal of the $p$-$d$ ligand contribution to the crystal field
splitting such that the $e_g$ states of the nominally Au$^{3+}$ cation
are energetically lower than the corresponding $t_{2g}$ states.
\end{abstract}

\maketitle

\section{Introduction}

The concept of an isolated ion interacting with a \emph{crystal field}
potential created by the surrounding ions, is perhaps one of the most
important and useful concepts in the theory of transition metal (TM)
oxides and materials containing rare-earth ions. Even though the
assumption of an isolated ion is of course an oversimplification,
crystal field theory has provided many important insights and is
crucial for our understanding of many materials with partially-filled
$d$ or $f$ electron
shells.\cite{Griffith:1961,Anderson:1963,Cox:1992,Figgis/Hitchman:book,Fazekas:1999,Bersuker:2010}

A prominent example illustrating the effect of the crystal field is
the case of a TM cation octahedrally coordinated by anions. Here, the
electro-static potential created by the negatively charged anions
splits the otherwise five-fold degenerate $d$ states of the TM cation
into the three-fold degenerate $t_{2g}$ and the doubly-degenerate
$e_g$ states. The remaining degeneracies can further be removed by
small distortions of the anionic octahedra which lead to additional
contributions to the crystal field.

The first principles calculation of crystal field splittings within a
solid or a molecular complex has a long history, perhaps starting with
the work of Van Vleck, who calculated the splitting between the $d$
electrons in the molecular cluster $X \cdot 6$H$_2$O (with $X$=Ti, V,
Cr) within a point charge model using hydrogenic
wave-functions.\cite{VanVleck:1939} A nice summary of the historical
developments following Van Vleck's work can be found in
Refs.~\onlinecite{Anderson:1963} and
\onlinecite{Figgis/Hitchman:book}. While initially only the
electro-static potential of the surrounding ions was considered as the
source of the observed level splittings, it soon became clear that
covalency effects, in particular hybridization between the electronic
orbitals of the central ion and the surrounding ligands, need to be
taken into account in order to arrive at a quantitatively correct
description.\cite{Sugano/Shulman:1963,Watson/Freeman:1964} As a
result, complex quantum chemical calculations for large molecular
clusters are required for accurate first principles calculation of
crystal field splittings, often termed {\it ligand field splitting} to
indicate that covalency effects are also taken into account.  In this
case, it is not clear a priori how many coordination shells have to be
included in the calculations or how to identify the relevant localized
orbitals and corresponding energy levels.

Recently, Wannier functions constructed from first-principles
electronic structure calculations have been used increasingly often to
represent atomic-like orbitals within a periodic crystal and to
calculate the corresponding level splittings (see
e.g. Refs.~\onlinecite{Pavarini_et_al:2005,Streltsov_et_al:2005,Solovyev_2:2006,Kovacik/Ederer:2010,Haverkort/Zwierzycki/Andersen:2012,Novak/Knizek/Kunes:2013}).
The use of Wannier functions is conceptually pleasing, since by
construction the Wannier functions form an orthonormal set of orbitals
that incorporates all effects of the crystal potential and provides a
complete basis set to represent the Bloch eigenstates of the system.
In addition, Wannier functions can often be constructed such that they
resemble atomic orbitals with a specific orbital character (e.g. $s$,
$p$, $d$, etc.), and thus allow for an intuitive interpretation of the
electronic structure within a tight-binding (TB) picture.  However,
the inherent non-uniqueness of the Wannier functions in principle
allows multiple ways for extracting TB parameters, leading to the
question of which Wannier functions are most appropriate for obtaining
energy splittings and hopping parameters.

Here, we use maximally localized Wannier functions to show how, by
constructing sets of Wannier functions corresponding to different
``energy windows'', one can successively isolate the effects of
hybridization with different ligand states.  This allows us to
distinguish, at least in an approximate way, between the pure
electro-static and the various covalent contributions to the crystal
field splitting for the $d$ orbitals of a TM cation.

The present article is structured in the following way: In
Sec.~\ref{MLWF} we discuss the general features of maximally localized
Wannier functions, explain our notation, and specify some details of
our calculations. In Sec.~\ref{sec_SrVO3} we introduce our method to
separate the electro-static and covalent contributions to the crystal
field splitting using the cubic perovskite SrVO$_3$ as example. We
then present results for two series of hypothetical tetragonally
distorted perovskite oxides in Sec.~\ref{sub_series} and
\ref{sub_hybEn}, and we discuss the relation of the calculated Coulomb
contribution to the crystal field and a simple electro-static point
charge model. Finally, in Sec.~\ref{sec:CAC} we apply our method to
the interesting case of the charge transfer insulator CsAuCl$_3$ and
analyze the different contributions to the crystal field splitting for
the Au$^{3+}$ cation. We show that the $p$-$d$ contribution to the
ligand field splitting is reversed relative to the $s$-$d$ and
electro-static contributions, which results in a lower energy of the
$e_g$ states compared to the $t_{2g}$ states. In
Appendix~\ref{App_MLWFCons} we demonstrate the consistency of our
choice of Wannier functions by considering the limit of large energy
separation between the nominal TM-$d$ and O-$p$ bands.

\section{Construction of maximally localized Wannier functions}
\label{MLWF}

The electronic states within a solid are usually represented in a
basis of Bloch states that correspond to eigenfunctions of the
single-particle Hamiltonian for electrons within the periodic
effective crystal potential.  In the following we will use density
functional theory, where this effective crystal potential is the
Kohn-Sham potential that consists of the electro-static Coulomb
potential created by the nuclei and of the Hartree and
exchange-correlation contributions arising from the interaction
between the electrons.\cite{Hohenberg/Kohn:1964,Kohn/Sham:1965} The
Bloch states are characterized by a wave-vector $\vec{k}$ within the
first Brillouin zone (BZ) and a band index $n$. The Hamiltonian is
diagonal in this basis:
\begin{equation}
\hat{H} = \sum_{n\vec{k}} \epsilon_{n\vec{k}}
\hat{a}^\dagger_{n\vec{k}} \hat{a}_{n\vec{k}} \quad .
\end{equation}
Here, $\hat{a}^\dagger_{n\vec{k}}$ is the creation operator for an
electron in the Bloch state $|\psi_{n\vec{k}}\rangle$, and
$\epsilon_{n\vec{k}}$ is the corresponding single particle energy.

The Bloch functions are transformed into a set of Wannier functions
$|w_{\alpha\vec{T}}\rangle$ in the following way (see
Ref.~\onlinecite{Marzari_et_al:2012}):
\begin{equation}
\label{eq:wf}
|w_{\alpha\vec{T}}\rangle = \frac{V}{(2\pi)^3} \int_\text{BZ} d^3k
e^{-i\vec{k}\vec{T}} \sum_n U^{(\vec{k})}_{n\alpha} | \psi_{n\vec{k}}
\rangle \quad .
\end{equation}
The Wannier orbitals are characterized by a unit cell index $\vec{T}$
and an additional index $\alpha$ which distinguishes different Wannier
orbitals within the same unit cell. This index can for example
indicate orbital and spin character as well as a specific site within
the unit cell.  $U^{(\vec{k})}$ is an arbitrary $k$-dependent unitary
matrix that mixes the various Bloch functions at the same
$k$-point. Different choices for $U^{(\vec{k})}$ lead to different
Wannier orbitals, which are therefore not uniquely defined by
Eq.~(\ref{eq:wf}).

A possible way to define a unique set of Wannier functions is to
choose a ``gauge'' that minimizes the total quadratic spread $\Omega =
\sum_\alpha \left( \langle r^2 \rangle_\alpha - \langle \vec{r}
\rangle_\alpha^2 \right)$ of the Wannier orbitals to obtain so-called
\emph{maximally localized Wannier functions}
(MLWFs).\cite{Marzari/Vanderbilt:1997}

The Hamiltonian is not diagonal in the Wannier basis:
\begin{equation}
\hat{H} = \sum_{\alpha, \beta, \vec{T}, \vec{T}'}
h_{\alpha\vec{T},\beta\vec{T}'} \hat{c}^\dagger_{\alpha\vec{T}} \hat{c}_{\beta\vec{T}'}
\quad ;
\end{equation}
the corresponding matrix elements are given by:
\begin{equation}
\label{eq:tb}
h_{\alpha\vec{T},\beta\vec{T}'} = \frac{V}{(2\pi)^3} \int_\text{BZ}
d^3k e^{i\vec{k}(\vec{T}-\vec{T}')} \sum_n (U^{(\vec{k})}_{n\alpha})^*
\epsilon_{n\vec{k}} U^{(\vec{k})}_{n\beta} \quad .
\end{equation}

In the case of TM oxides, each MLWF is typically located on a specific
atomic site and has a clear dominant orbital character, which allows
interpretation of the \mbox{MLWFs} as TB orbitals. Thus, if either
$\vec{T} \neq \vec{T}'$ or the indices $\alpha$ and $\beta$ correspond
to orbitals $m$ and $m'$ at different sites $i$ and $j$, the matrix
elements, $h_{\alpha\vec{T},\beta\vec{T}'}$, in Eq.~(\ref{eq:tb}) can
be interpreted as the \emph{hopping amplitude} between $m i \vec{T}$
and $m' j \vec{T}'$ denoted by $t_{mi\vec{T},m'j\vec{T}'}$.  On the
other hand if $\alpha$ and $\beta$ correspond to the same site $i$ and
the same orbital $m$ in the same unit cell $\vec{T}=\vec{T}'$, then
the corresponding matrix elements, $\,h_{\alpha\vec{T},\beta\vec{T}}$,
represent the \emph{on-site energies}, $\varepsilon_{mi}$, of the TB
basis.  The differences in the on-site energies between orbitals with
the same predominant orbital character (e.g. TM-$d$ or O-$p$) can then
be interpreted as crystal field splittings. In what follows, the index
$m$ will take the values $x^2$, $z^2$, $xy$, $xz$, and $yz$ to
indicate Wannier functions with $x^2-y^2$ , $3z^2-r^2$, $xy$, $xz$,
and $yz$ orbital character, respectively. Furthermore, for the cubic
case $m$ can have the label $e_g$ or $t_{2g}$ indicating
Wannier functions with $e_g$ or $t_{2g}$ character, respectively.

We note that different sets of Wannier functions can be constructed
corresponding to different sets of bands, depending on which
bands $n$ are included in the summation in Eq.~(\ref{eq:wf}). As we
demonstrate in the next section, this feature can be used to construct
different TB models, corresponding to: i) a TB picture with only
(effective) TM-$d$ states, ii) a so-called ``$p$-$d$ model''
containing both TM-$d$ and O-$p$ orbitals, or iii) other TB models
containing either more or fewer basis states (see
e.g. Refs.~\onlinecite{Kovacik/Ederer:2011},
\onlinecite{Lechermann_et_al:2006}, or
\onlinecite{Aichhorn_et_al:2009}).

All electronic-structure calculations presented in this work are
performed with the ``Vienna Ab-initio Simulation package'' ({\tt
  VASP}) using projector-augmented wave
potentials.\cite{Kresse/Furthmueller_CMS:1996,Kresse/Joubert:1999}
MLWFs are then constructed by employing the {\tt vasp2wannier90}
interface\cite{Franchini_et_al:2012} in combination with the {\tt
  wannier90} code.\cite{Mostofi_et_al:2008} All calculations are
performed within the generalized gradient approximation according to
Perdew, Burke, and Ernzerhof for the non spin-polarized case.\cite{Perdew/Burke/Ernzerhof:1996}. The calculations for Sr$M$O$_3$ and Tb$M$O$_3$ presented in
Secs.~\ref{sec_SrVO3} and \ref{sub_series} are performed using a
$\Gamma$-centered $6 \times 6 \times 6$ mesh for k-point sampling and
a plane wave kinetic energy cutoff of 500\,eV, whereas the
calculations for CsAuCl$_3$ presented in Sec.~\ref{sec:CAC} are
performed using a $\Gamma$-centered $5 \times 5 \times 5$ mesh and a
cutoff energy of 350\,eV. We include both 3$s$ and 3$p$ semi-core
states in the valence for all 3$d$ TM elements from Sc to Fe, while
for the elements from Co to Zn only the 3$p$ states are included. 4$s$
and 4$p$ states are treated as valence states for Sr, whereas no
semi-core states are included in the valence for Au. For the Tb
pseudopotential 4$f$ states are ``frozen'' in the core while 5$p$
states are treated as valence states.

\section{Results and Discussion}
\label{Results}

\subsection{Coulomb and hybridization contributions to the $e_g$-$t_{2g}$ splitting in SrVO$_3$}
\label{sec_SrVO3}

We first demonstrate our approach for the simple case of SrVO$_3$,
which crystallizes in the ideal cubic perovskite structure and
exhibits a band structure with well-separated groups of bands
corresponding to different dominant orbital characters.

\begin{figure}
\includegraphics[width=\columnwidth]{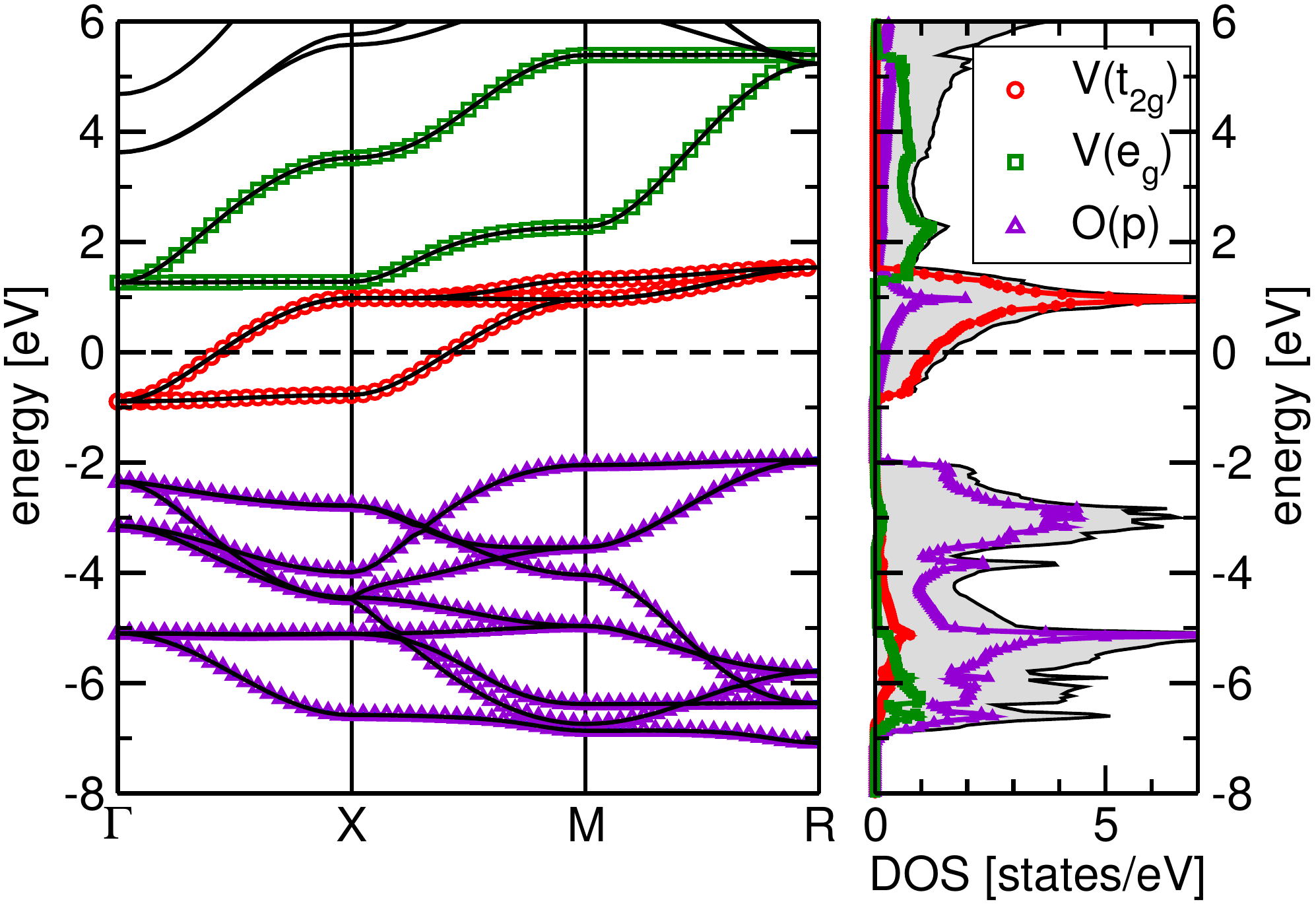}
\caption{(Color online) Left: calculated band structure of SrVO$_3$
  along the high-symmetry lines of the simple cubic Brillouin zone
  (thin black lines). The different symbols (colors) correspond to the
  dispersion calculated from the MLWFs calculated separately for the
  three groups of bands (see main text).  Right: orbital- and
  site-projected density of states (DOS) for SrVO$_3$. Different
  symbols (colors) correspond to projections on different atomic
  orbitals as indicated in the legend. The total DOS is represented by
  the dark shaded region. The Fermi-level is set to zero energy.}
\label{fig:SVO-bands}
\end{figure}

Fig.~\ref{fig:SVO-bands} shows the Kohn-Sham band structure obtained
for SrVO$_3$. Three groups of bands can be recognized in the relevant
energy region around the Fermi level, which are highlighted by the
different symbols in Fig.~\ref{fig:SVO-bands}. A comparison with the
orbital- and site-projected density of states shows that these three
groups correspond to bands with predominant O-$p$, V-$d$($t_{2g}$),
and V-$d$($e_g$) orbital character, respectively. However, it can be
seen that the bands with predominant O-$p$ character also contain a
noticeable amount of V-$d$ character and vice versa.  This is a
typical situation found in many TM oxides and is indicative of the
partially covalent character of the TM-O bond.  For the present (and
more frequently observed) case where the O-$p$ states are
energetically lower than the TM-$d$ states, the bands with
predominant O-$p$ character represent the \emph{bonding} linear
combination of atomic orbitals (within a TB picture), whereas the
bands with predominant V-$d$ character represent the corresponding
\emph{anti-bonding} linear combination.

\begin{figure}
\begin{center}
\includegraphics[width=\columnwidth]{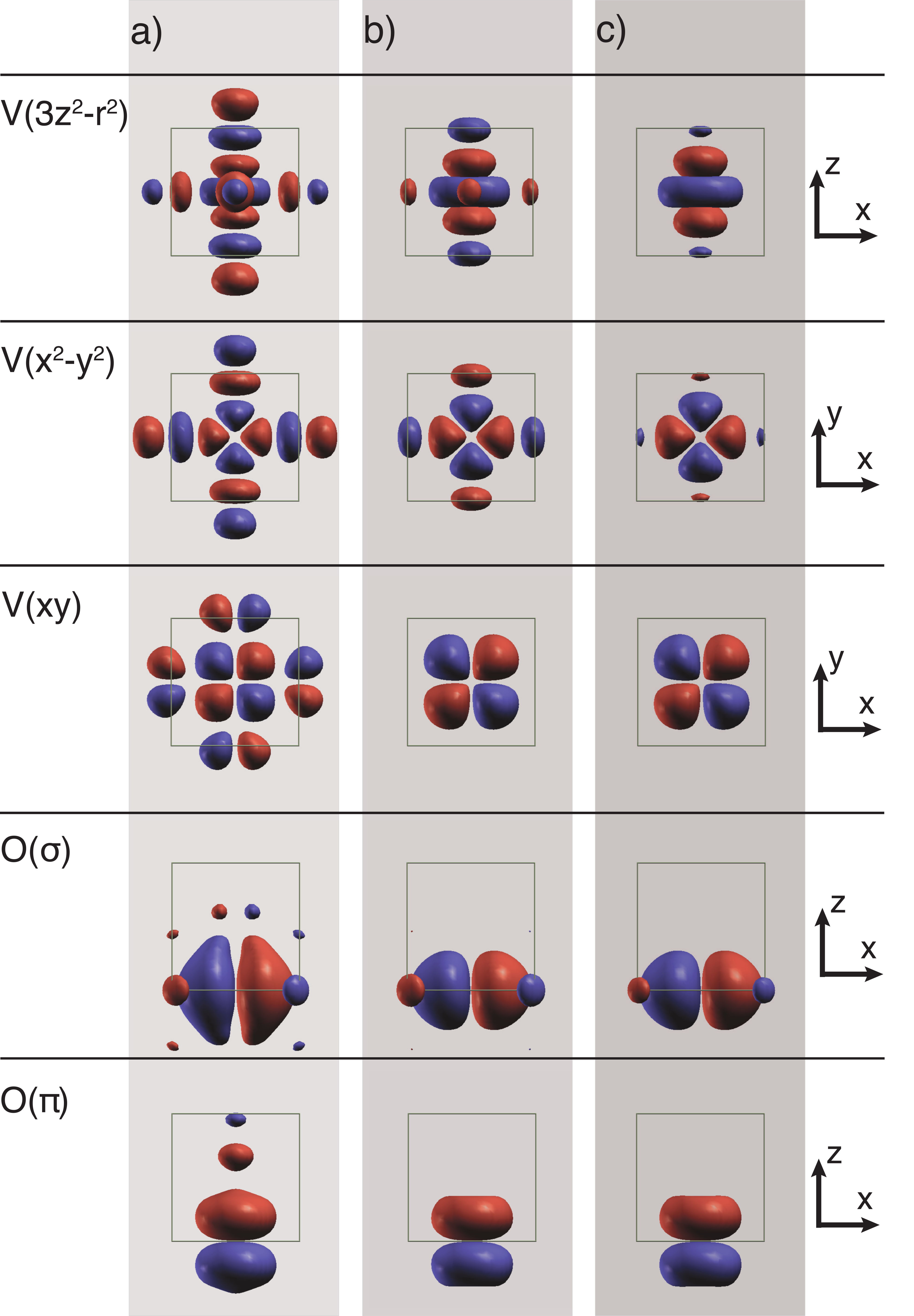}
\end{center}
\caption{(Color online) Representative members from three different
  sets of MLWFs. Column (a) corresponds to the MLWFs which are
  constructed \emph{separately} for each of the three groups of bands
  in Fig.~\ref{fig:SVO-bands}, whereas column (b) corresponds to the
  case where MLWFs are constructed \emph{simultaneously} for all of
  these bands. Column (c) contains Wannier functions constructed for
  the largest energy window containing also energetically lower-lying
  bands with predominant O-$s$ and Sr-$p$ (semi-core states)
  character. The first and second rows show the V-$e_g$-like Wannier
  functions, the third row shows one of the three equivalent
  V-$t_{2g}$-like Wannier functions, the fourth and fifth row show
  $\pi$- and $\sigma$-type oxygen $p$-like Wannier functions,
  respectively. The spatial orientation for each row is indicated on
  the right side of the figure.}
\label{fig:mlwfs}
\end{figure}

We now construct MLWFs \emph{separately} for each group of bands in
Fig.~\ref{fig:SVO-bands}. The resulting MLWFs are depicted in
Fig.~\ref{fig:mlwfs}(a). Each Wannier function is centered either on
an O or a V site and has an ``inner'' part which resembles the
dominant orbital character in the corresponding bands and an ``outer''
part that resembles atomic orbitals on the neighboring ions which
hybridize with the central orbital.  It can be seen that MLWFs with a
central V-$d$-like part correspond to anti-bonding combinations of
V-$d$ and O-$p$ atomic-like orbitals, whereas MLWFs with a central
O-$p$-like part represent bonding combinations of such orbitals.

The MLWFs constructed in this way can thus be viewed as ``molecular
orbitals'' that arise from hybridization between neighboring ions.
Combining molecular orbitals of the same ``type'' but within different
unit cells (by Bloch summation) then gives rise to the different
groups of bands in Fig.~\ref{fig:SVO-bands}.

We denote the on-site energies (see Sec.~\ref{MLWF}) for the
V-centered Wannier functions corresponding to this set of orbitals as
$\varepsilon^{(d)}_{m}$, where the superscript indicates that these
Wannier functions were obtained separately for the ``effective'' V-$d$
bands and the subscript labels different $d$ orbitals. These on-site
energies include both the effect of hybridization with the surrounding
ligands as well as the purely electro-static contribution to the
crystal potential. Consequently, the difference
$\varepsilon^{(d)}_{e_g} - \varepsilon^{(d)}_{t_{2g}}$, i.e. the
difference between on-site energies of the V-$e_g$-like and the
V-$t_{2g}$-like Wannier functions, can then be interpreted as the full
crystal plus ligand field splitting between the V-$d$ states in cubic
SrVO$_3$.

We then construct a second set of Wannier functions where all three
groups of bands in Fig.~\ref{fig:SVO-bands} are included in the
summation over $n$ in Eq.~(\ref{eq:wf}), i.e. we construct MLWFs
\emph{simultaneously} for all O-$p$- and V-$d$-like bands around the
Fermi level.  The resulting Wannier functions are shown in
Fig.~\ref{fig:mlwfs}(b). Again, all Wannier functions are centered on
either O or V sites, but now the V-$d$-like Wannier functions contain
only minimal contributions of O-$p$ orbitals at the surrounding
ligands and vice versa.  In analogy with the previous case, we denote
the on-site energies of the so-obtained V-$d$-like Wannier functions
as $\varepsilon^{(dp)}_{m}$ (V(d) and O(p) bands included). 
It can be seen from Fig.~\ref{fig:mlwfs}
that the inclusion of all nominal O-$p$ and V-$d$ bands in the Wannier
construction has essentially removed the (V-$d$)-(O-$p$) hybridization
from the resulting Wannier orbitals. This is consistent with the
maximum localization condition, since the spatial extension of the
Wannier functions can be reduced by distributing the O-$p$ and V-$d$
character contained in the Bloch functions of the corresponding bands
over different Wannier functions.

One can still recognize a rather strong O-$s$ admixture in the
V-$e_g$-like Wannier functions and some ``tails'' in the
$\pi$-oriented O-$p$ Wannier functions located at the surrounding Sr
ions, but overall these Wannier functions are much more similar to
atomic orbitals than the Wannier functions of the first set.  We note
that the ``size'' of the admixtures that are visible in the pictures
shown in Fig.~\ref{fig:mlwfs} depends entirely on the iso-surface
value chosen for visualization of the Wannier functions.

\begin{table}
\caption{Splitting between $e_g$ and $t_{2g}$-type Wannier functions
  for different sets of MLWFs corresponding to different energy
  windows as described in the main text.}
\label{tab:SVO-split}
\begin{ruledtabular}
\begin{tabular}{ccc}
\begin{minipage}{0.5\columnwidth}
Bands included in the\\
Wannier construction
\end{minipage}
& Symbol & Value [eV] \\[7pt]
\hline
only (effective) Mn-$d$  & 
$\varepsilon^{(d)}_{e_g} - \varepsilon^{(d)}_{t_{2g}}$ &  2.69 \\
Mn-$d$ and O-$p$ & 
$\varepsilon^{(dp)}_{e_g} - \varepsilon^{(dp)}_{t_{2g}} $ & 1.69 \\
Mn-$d$, O-$p$, and O-$s$ & 
$\varepsilon^{(dps)}_{e_g} - \varepsilon^{(dps)}_{t_{2g}}$ &1.17 \\
Mn-$d$, O-$p$, O-$s$, Sr-4$p$ & 
$\varepsilon^{(dpsp)}_{e_g} - \varepsilon^{(dpsp)}_{t_{2g}} $ & 1.14 \\
\end{tabular}
\end{ruledtabular}
\end{table}

If one compares the splitting of the on-site energies between
V-$e_g$-like and V-$t_{2g}$-like Wannier functions for the two sets
(see Table~\ref{tab:SVO-split}), one can see that this splitting is
reduced by 1\,eV in the second set compared to the first.  This 1\,eV
reduction can therefore be viewed as the contribution to the splitting
stemming from hybridization of the central V-$d$ orbitals with the $p$
orbitals of the surrounding oxygen ligands.

We now construct two more sets of Wannier orbitals by including
additional bands at lower energies (not shown in
Fig.~\ref{fig:SVO-bands}) in the Wannier construction. First we include
bands with predominant O-$s$ character (located around 19\,eV below
the Fermi level) and then also lower-lying bands with Sr-$p$ character
(semi-core states; around 16\,eV below the Fermi level). In the first
case we denote the on-site energies of the V-centered Wannier
functions as $\varepsilon^{(dps)}_{m}$ (V-$d$, O-$p$, and O-$s$
included), whereas in the second case we denote them as
$\varepsilon^{(dpsp)}_{m}$ (V-$d$, O-$p$, O-$s$, and Sr-$p$
included). Some representative Wannier functions of this latter set
are shown in Fig.~\ref{fig:mlwfs}(c). It can be seen that these
Wannier functions contain only a minimal amount of mixing between
atomic-like orbitals on different sites.  A certain admixture is
necessary, however, to ensure orthonormality between the Wannier
functions.

The V-$d$ crystal field splittings corresponding to these two new sets
of Wannier functions are also included in Table~\ref{tab:SVO-split}.
It can be seen that removing (or minimizing) the hybridization between
V-$e_g$ and O-$s$ orbitals in the ``$dps$'' Wannier functions further
reduces the $e_g$-$t_{2g}$ crystal field splitting by $\sim$0.5\,eV
compared to the set where only the nominal V-$d$ and O-$p$ bands are
included in the construction of the Wannier functions. On the other
hand, the inclusion of the Sr-$p$ semi-core states in the Wannier
construction has only a negligible effect since these states do not
hybridize strongly with the V-$d$ orbitals.

We conclude, therefore, that by including more and more bands in the
construction of the Wannier functions, i.e. by successively removing
the effect of inter-site hybridization on the resulting orbitals, the
splitting between the $e_g$- and $t_{2g}$-like Wannier functions
converges to a value which can be viewed as the purely electro-static
contribution to the crystal field splitting.  The difference of this
value compared to the calculated splittings for Wannier functions that
include hybridization with specific ligand orbitals then allows
quantification of the various hybridization contributions to the
crystal field splitting.  In the present case, this means that the
total $e_g$-$t_{2g}$ splitting in SrVO$_3$ of 2.69\,eV contains a
contribution of $\sim$1.14\,eV of electro-static origin, a
contribution of about 0.5\,eV originating from $d$-$s$ hybridization,
and 1\,eV originating from $d$-$p$ hybridization.

Before extending our analysis of the above-defined ``electro-static''
part of the crystal field splitting to the case of
tetragonally-distorted perovskite, we comment on another method
sometimes used to estimate such a contribution.  This method is based
on the fact that in a simple TB model for TM perovskite, which only
includes O-$p$ and TM-$d$ orbitals and hopping between O and TM
nearest neighbors, the energy of the $d$-bands at the $\Gamma$-point
is equal to the on-site energy of the respective orbitals.  Thus, the
$\Gamma$-point splitting observed in the band structure calculation is
often taken as representative of the electro-static part of the
crystal field splitting.  We note that this interpretation, however,
neglects the effects induced by hopping between TM $e_g$-like and O
$s$-like Wannier functions ($\sim 0.5$\,eV, as shown in
Table~\ref{tab:SVO-split}) as well as further neighbor hoppings which,
as in the case of SrVO$_3$, can be quite substantial. Indeed, the
splitting between nominal $e_g$ and $t_{2g}$ bands at the
$\Gamma$-point (2.16\,eV for SrVO$_3$, see Fig.~\ref{fig:SVO-bands})
is not only significantly different from the splitting
$\varepsilon^{(dps)}_{e_g} - \varepsilon^{(dps)}_{t_{2g}}$ (1.17 eV),
but is also quite different from the value $\varepsilon^{(dp)}_{e_g} -
\varepsilon^{(dp)}_{t_{2g}}$ (1.69 eV) which includes the effect of
hybridization between $e_g$-like and $s$-like Wannier functions. The
discrepancy with the latter value, for the case of SrVO$_3$, can be
explained by the non-negligible amplitude of the hopping between
neighboring TM $t_{2g}$-like Wannier functions ($\sim -0.11$ eV)
compared to that between the corresponding TM $e_{g}$-like Wannier
functions (approximately one order of magnitude smaller).  At the
$\Gamma$-point, the presence of these hoppings shifts the $t_{2g}$
bands down from their on-site energy while leaving the $e_g$ bands at
a value similar to the $e_g$ on-site energy.  The down-shift in the TM
$t_{2g}$ bands at $\Gamma$ is approximately $0.44$ eV (there are four
neighbors with a hopping of $-0.11$ eV per $t_{2g}$ Wannier function)
which explains the apparent discrepancy between the on-site energy
splittings and the splittings of the bands at the $\Gamma$-point.

\subsection{Tetragonal crystal field splitting in the series Sr$M$O$_3$ and Tb$M$O$_3$}
\label{sub_series}

Next we extend our analysis to the case of tetragonally distorted
perovskite systems, and focus particularly on the ability of our
method to correctly extract the smaller splittings occurring
\emph{within} the $t_{2g}$ and $e_g$ manifolds. We consider the series
of compositions Sr$M^{4+}$O$_3$ and Tb$M^{3+}$O$_3$ with $M$ being a
3$d$ TM cation. To be able to systematically compare the results for
each $M$, all calculations are performed for a hypothetical
tetragonally distorted perovskite structure (space group $P4/mmm$)
with identical lattice parameters for each composition
($a=3.86$\,\AA{} and $c/a=1.1$). We focus on the crystal field
splitting of the ``$dps$'' basis set, where admixtures between the
TM-$d$ states and the surrounding ligand $s$ and $p$ states are
minimized in the Wannier functions.

>From a simple electro-static model, one would expect that for a
tetragonally distorted octahedron with $c/a>1$ the $x^2-y^2$-type
orbital should have higher energy than the $3z^2-r^2$-type orbital,
due to the larger lattice constant along $c$ (the distance to the
negatively charged ligands is larger along $z$ than along $x$ and $y$,
leading to a reduction of the Coulomb energy of $3z^2-r^2$ compared to
$x^2-y^2$).  Similarly, the $xy$-type orbital is expected to have
higher energy than the $xz$- and $yz$-type orbitals.  We therefore
define energy splittings $\Delta_{e_g}=\varepsilon^{(dps)}_{x^2}
-\varepsilon^{(dps)}_{z^2}$ and
$\Delta_{t_{2g}}=\varepsilon^{(dps)}_{xy}-\varepsilon^{(dps)}_{xz/yz}$.
 \footnote{Although the $t_{2g}$ and $e_{g}$ notation is only strictly
   valid for cubic systems, here, we use these symbols to refer,
   respectively, to the three low-lying and to the two high-lying 3$d$
   energy levels of an ion in a ligand field slightly distorted from a
   cubic one.}  For the present case ($c/a>1$), the simple
 electro-static model would predict these energy splittings to be
 positive.
\begin{figure}
\includegraphics[width=0.8\columnwidth]{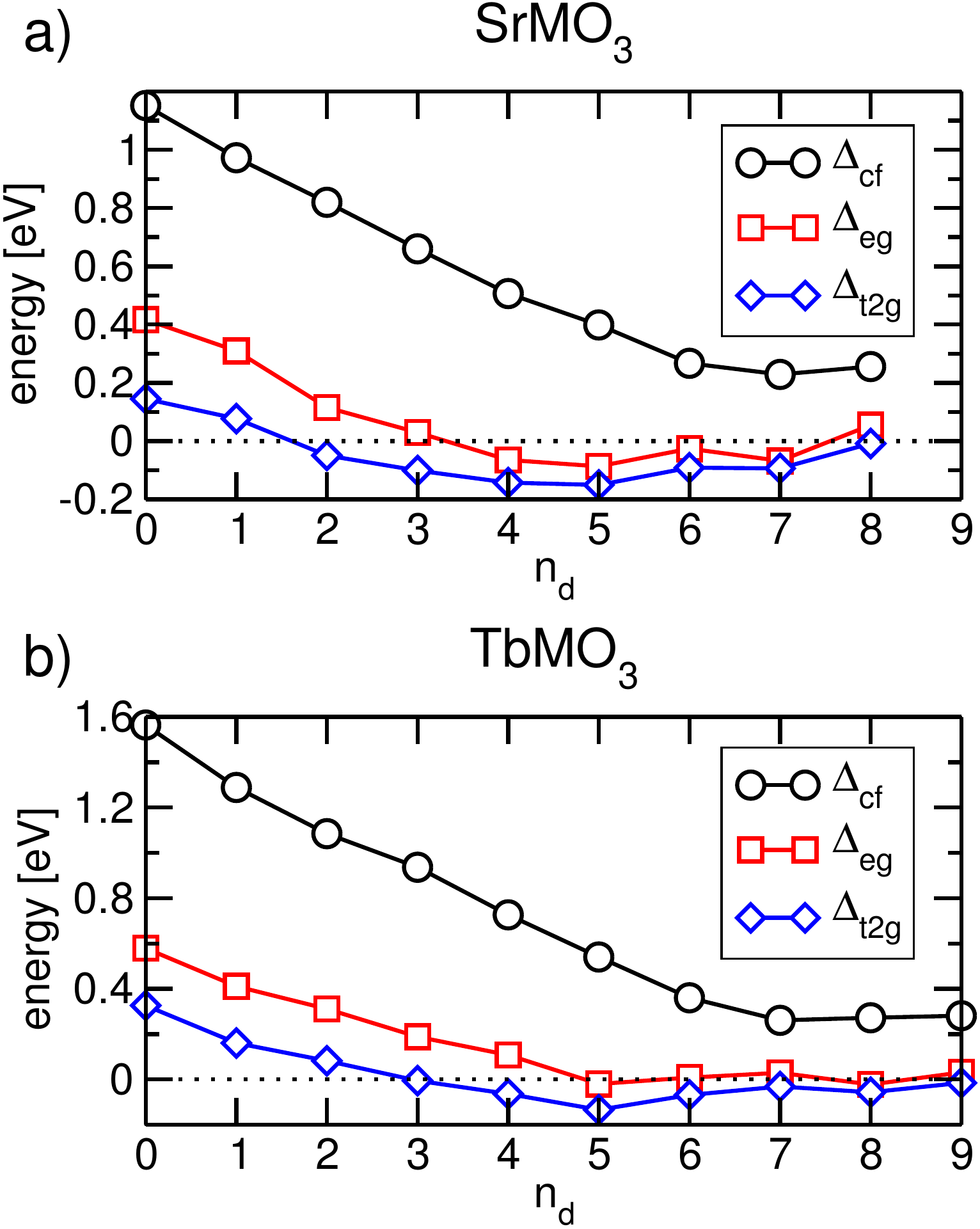}
\caption{(Color online) Calculated crystal field splittings (see main
  text) for both Sr$M$O$_3$ and Tb$M$O$_3$ as a function of the formal
  $d$ electron count $n_d$. For the case of Sr$M$O$_3$ (Tb$M$O$_3$),
  $n_d=0$ corresponds to SrTiO$_3$ (TbScO$_3$) and $n_d=8$ corresponds
  to SrZnO$_3$ (TbCuO$_3$).}
\label{fig:series}
\end{figure}

Figs.~\ref{fig:series}(a) and (b) show, respectively, the calculated
energy splittings for both Sr$M$O$_3$ and Tb$M$O$_3$ as a function of
the formal $d$ electron occupation $n_d$. Here, $\Delta_\text{cf}$ is
the energy difference between the average on-site energy of the two
$e_g$ orbitals and that of the three $t_{2g}$ orbitals. It can be seen
that there is an overall decrease of the absolute value of the crystal
field splittings with increasing $n_d$, and that the crystal field is
somewhat stronger in the Tb$M$O$_3$ series compared to the Sr$M$O$_3$
compounds. These trends can be explained by changes in the spread,
i.e.  the spatial extension of the Wannier functions, which decreases
with increasing $n_d$ and is larger for the TM cations with lower
valence. Thus, the more localized the Wannier functions, the weaker
its interactions with the surrounding crystal field.

Furthermore, it can be seen from Fig.~\ref{fig:series} that
$\Delta_\text{cf}$ has the expected positive sign for all compounds,
i.e. the $e_g$ orbitals are on average higher in energy than the
$t_{2g}$ orbitals. In contrast, the much smaller tetragonal components
of the crystal field splitting, $\Delta_{e_g}$ and $\Delta_{t_{2g}}$,
change sign across the series.  For a small number of $d$ electrons
($n_d < 2$ and $n_d < 3$, respectively, for the Sr$M$O$_3$ and
Tb$M$O$_3$ cases), both $\Delta_{e_g}$ and $\Delta_{t_{2g}}$ are
positive, as expected from the simple electro-static model.  However,
both splittings become negative for increasing $d$ electron
count. Towards the end of the series the trend is reversed, and a
positive sign of $\Delta_{e_g}$ is observed (for $n_d=8$ and $n_d=9$,
corresponding to SrZnO$_3$ and TbZnO$_3$, respectively).  In the
following we discuss several possible reasons for the sign change of
the calculated tetragonal crystal field splittings and for the
disagreement between these splittings and the expected
``electro-static contribution'' to the crystal field based on a simple
point charge model.

A first point to note is that, as already mentioned, the spread of the
Wannier functions depends both on the number of $d$ electrons (or,
equivalently, the nuclear charge) as well as on the valence state of
the corresponding TM cation. In addition, the spread is different for
Wannier functions which are not symmetry equivalent to each other
within tetragonal symmetry, e.g. the $x^2-y^2$-like Wannier functions
are slightly more localized than the $3z^2-r^2$-like Wannier functions
and thus experience a slightly weaker effective crystal field
potential. However, it seems unlikely that these small differences
will indeed change the sign of the crystal field splitting in an
appropriately modified point charge model.

We also point out that even the Wannier functions corresponding to the
largest energy window, i.e. with minimal ligand hybridization, are not
completely identical to simple atomic orbitals. The Wannier functions
cannot be separated into angular and radial parts, and although they
have a dominant orbital character, they might not be completely
appropriate for a comparison to the crystal field splitting expected
for atomic orbitals surrounded by point charges. The Wannier functions
must also contain ``tails'' localized at the surrounding ligand sites
to ensure orthogonality between the Wannier functions located at
different sites. Interestingly, it was shown by Kleiner that
orthogonalization of the atomic orbitals with respect to the ligand
orbitals is indeed necessary to obtain a realistic crystal field
splitting.\cite{Kleiner:1952} Otherwise the positively charged nuclei
of the ligand atoms are too attractive for the $d$ electrons of the
central TM ion. However, such orthogonalization also makes it
conceptually more difficult to clearly distinguish between
electro-static and hybridization effects. It is therefore unclear how
these tails will influence the calculated crystal field splittings.

One could also ask whether the maximum localization condition is
necessarily the best way to define unique Wannier functions for the
evaluation of crystal field splittings, or whether other possibilities
based on atomic projections \cite{Ku_et_al:2002} or symmetry-based
criteria \cite{Sakuma:2013} might be more appropriate. While a
detailed analysis of this point is beyond the scope of this article,
we show in Appendix~\ref{App_MLWFCons} that, at least for the present
case of tetragonally distorted perovskite systems, MLWFs provide a
consistent description and even appear to be somewhat preferable to
Wannier functions based on atomic projections.

Finally, we note that the crystal field energies calculated in this
work correspond to the energies of the Wannier orbitals within the
Kohn-Sham potential. However, apart from the electro-static potential
created by the surrounding charges, the Kohn-Sham potential also
contains the electro-static interaction with the other electrons on
the same atom plus contributions due to exchange and correlation. It
is unclear how this will affect the level splittings. The fact that
for small $n_d$ the calculated splittings have the expected sign,
while for larger $n_d$ the sign changes, seems to indicate that the
interaction with the other $d$ electrons on the same site could indeed
play an important role here.

We point out that it is unclear how an ``exact'' calculation of the
purely electro-static contribution to the crystal field splitting
should be carried out even in principle. One possibility would
be to divide the electronic charge density according to contributions
corresponding to the different occupied (or partially
occupied) Wannier functions, and use this decomposition of the charge
density to distinguish between the electro-static potential generated
by electrons on the same site and on the surrounding
sites. Calculation of the electro-static contribution to the crystal
field would then amount to calculating the energy of the on-site
Wannier functions in the electro-static potential generated by the
surrounding charge (electrons sitting in different Wannier centers
located at different sites plus the charge of the corresponding ionic
cores).

However, while such a calculation is in principle possible, it is
unclear whether further insights can be gained by such an elaborate
procedure. In the present work, we therefore use the method described
in Sec.~\ref{sec_SrVO3} as working definition for the separation of
the electro-static contribution and the different hybridization
contributions to the ligand field splitting, and we show that valuable
insight can be obtained from this decomposition. To this end, in the
following section we continue our discussion on tetragonally distorted
perovskites, and in Sec.~\ref{sec:CAC} we use our method to
demonstrate the competition between electro-static and certain
hybridization contributions to the ligand field splitting in a
negative charge transfer system.

\subsection{Hybridization energies across the Sr$M$O$_3$ and Tb$M$O$_3$ series}
\label{sub_hybEn}

\begin{figure}
\includegraphics[width=1\columnwidth]{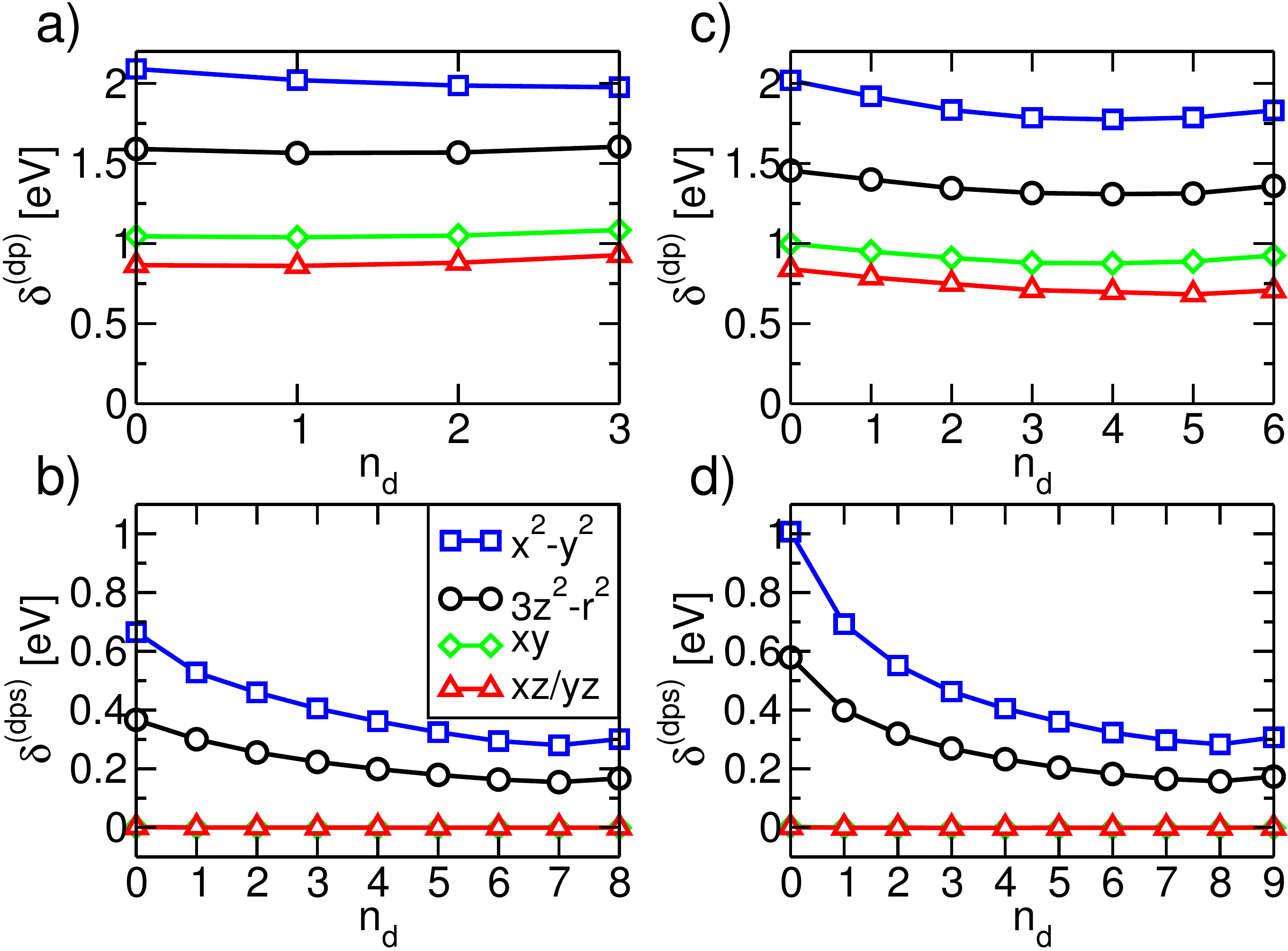}
\caption{(Color online) Estimated hybridization energies
  $\delta^{(dp)}_{m}$, panels a) and c), and $\delta^{(dps)}_{m}$, panels
  b) and d), as defined in the text. Panels a) and b) show the $n_d$
  dependence of the hybridization energies for the Sr$M$O$_3$ series,
  while panels c) and d) show the corresponding data for the
  Tb$M$O$_3$ series. Different symbols/colors indicate the orbital
  character of the different Wannier functions as defined in the inset
  of panel b).}
\label{fig:hybrid}
\end{figure}

We now discuss the strength of the $p$-$d$ hybridization across the
two series of tetragonally distorted perovskite systems. As a quantity
to represent the strength of the TM-$d$-O-$p$ hybridization we
consider the difference between the on-site energies of the Wannier
functions obtained by including only the nominal TM-$d$ bands and
those obtained by also including O-$p$ bands,
i.e. \mbox{$\delta^{(dp)}_m = \varepsilon^{(d)}_{m} -
  \varepsilon^{(dp)}_{m}$} (where $m$ indicates the orbital character
of the corresponding Wannier functions). This quantity (more precisely
$\varepsilon^{(d)}_m$) is only well defined for cases where the bands
with predominant TM-$d$ character are not entangled with the bands
with predominant O-$p$ character, i.e. the corresponding energies are
well separated. In our calculations for the Sr$M$O$_3$ and Tb$M$O$_3$
series this is the case for $0 \leq n_d \leq 3$ (Sr$M$O$_3$) and $0
\leq n_d \leq 6$ (Tb$M$O$_3$). Figs.~\ref{fig:hybrid}(a) and (c) show
the $n_d$ dependence of $\delta^{(dp)}_m$ for the five $d$-like WFs
for Sr$M$O$_3$ and Tb$M$O$_3$, respectively.

The hybridization is stronger for TM-$e_g$-like Wannier functions than
for the TM-$t_{2g}$-like Wannier functions. This is due to the
stronger $\sigma$-bonding between the $e_g$ orbitals with the
neighboring O-$p$ orbitals compared to the $\pi$-bonding between
TM-$t_{2g}$ and O-$p$. Moreover, $\delta^{(dp)}_{x^2} >
\delta^{(dp)}_{z^2}$ and $\delta^{(dp)}_{xy} > \delta^{(dp)}_{xz/yz}$
for all the $n_d$ values, as one would expect for the considered
tetragonal distortion with $c/a > 1$. We also notice that $
\delta^{(dp)}_{m}$ is more or less constant throughout the series and
quite similar for the Tb- and Sr-based compounds.

Similarly, to estimate the strength of hybridization between
TM-$d$-like and O-$s$-like Wannier functions we define
$\delta^{(dps)}_m = \varepsilon^{(dp)}_{m} -
\varepsilon^{(dps)}_{m}$. The dependence of this quantity on $n_d$ for
Sr$M$O$_3$ and Tb$M$O$_3$ is plotted, respectively, in
Fig.~\ref{fig:hybrid} b) and d). For these hybridization energies we
find that $\delta^{(dps)}_{t_{2g}} \approx 0$ and
$\delta^{(dps)}_{x^2} >\delta^{(dps)}_{z^2}$.  This behavior can be
explained by the lack of TM-$t_{2g}$-O-$s$ hybridization (due to
symmetry reasons) and by the tetragonal distortion, which increases
the overlap between TM-$x^2$-like Wannier functions with O-$s$-like
Wannier functions compared to that of TM-$z^2$-like Wannier functions.

Furthermore, we notice that the $n_d$ dependence of the TM-$e_g$-O-$s$
hybridization is more pronounced than that of the TM-$d$-O-$p$
hybridization. The reason for this is
the fact that the O-$p$-bands are much closer in energy to the TM-$d$-bands than the
O-$s$-bands. Indeed, qualitatively, the strengths of the TM-$d$-O-$p$
and TM-$d$-O-$s$ hybridizations are proportional to
$t^2_{d,p}/(\varepsilon_d - \varepsilon_p )$ and
$t^2_{d,s}/(\varepsilon_d - \varepsilon_s )$, respectively, where $t^2_{d,p(s)}$ is the
TM-$d$-O-$p$ (TM-$d$-O-$s$) hopping and $\varepsilon_i$ is the on-site
energy of the $i$-th Wannier function ($i=s,p,d$).(see also
Appendix~\ref{App_MLWFCons}) We find that
$t^2_{d,p}$ and $t^2_{d,s}$ decay strongly with increasing $n_d$. The
splitting $\varepsilon_d - \varepsilon_p$ also decreases strongly over
the series (the TM-$d$ and O-$p$ bands overlap for $n_d > 3$ and $n_d
> 6$, respectively, for SrMO$_3$ and TbMO$_3$), which partly compensates the
decay in $t^2_{d,p}$ yielding the almost constant behavior of
$\delta^{(dp)}_{d}$ (see Fig.~\ref{fig:hybrid} a) and c)).  However,
although the absolute variation in $\varepsilon_d - \varepsilon_s$ as
a function of $n_d$ is similar to that of $\varepsilon_d -
\varepsilon_p$, its relative variation is small due to its larger
value. This leads to a functional behavior of
$t^2_{d,s}/(\varepsilon_d - \varepsilon_s )$ similar to that of
$t^2_{d,s}$ and clarifies the strong changes of $\delta^{(dps)}_{e_g}$
(see Fig.~\ref{fig:hybrid} b) and d)).

\subsection{The negative charge transfer system CsAuCl$_3$}
\label{sec:CAC}

In the following we apply our approach to the interesting case of
CsAuCl$_3$, demonstrating that useful insights can be gained by the
analysis of the different contributions to the crystal field splitting
and confirming the classification of CsAuCl$_3$ as a negative charge
transfer system with competing tendencies for the level ordering.

The average formal valence of Au in this material is $2+$,
corresponding to a valence electron configuration of 5$d^9$. However,
since such a $d^9$ configuration is unfavorable, there is a charge
disproportionation of the Au cation into Au$^{3+}$ and Au$^{1+}$ with
formal 5$d$ electron configurations of $d^{8}$ and $d^{10}$,
respectively.

\begin{figure}
\centerline{\includegraphics[width=0.5\columnwidth]{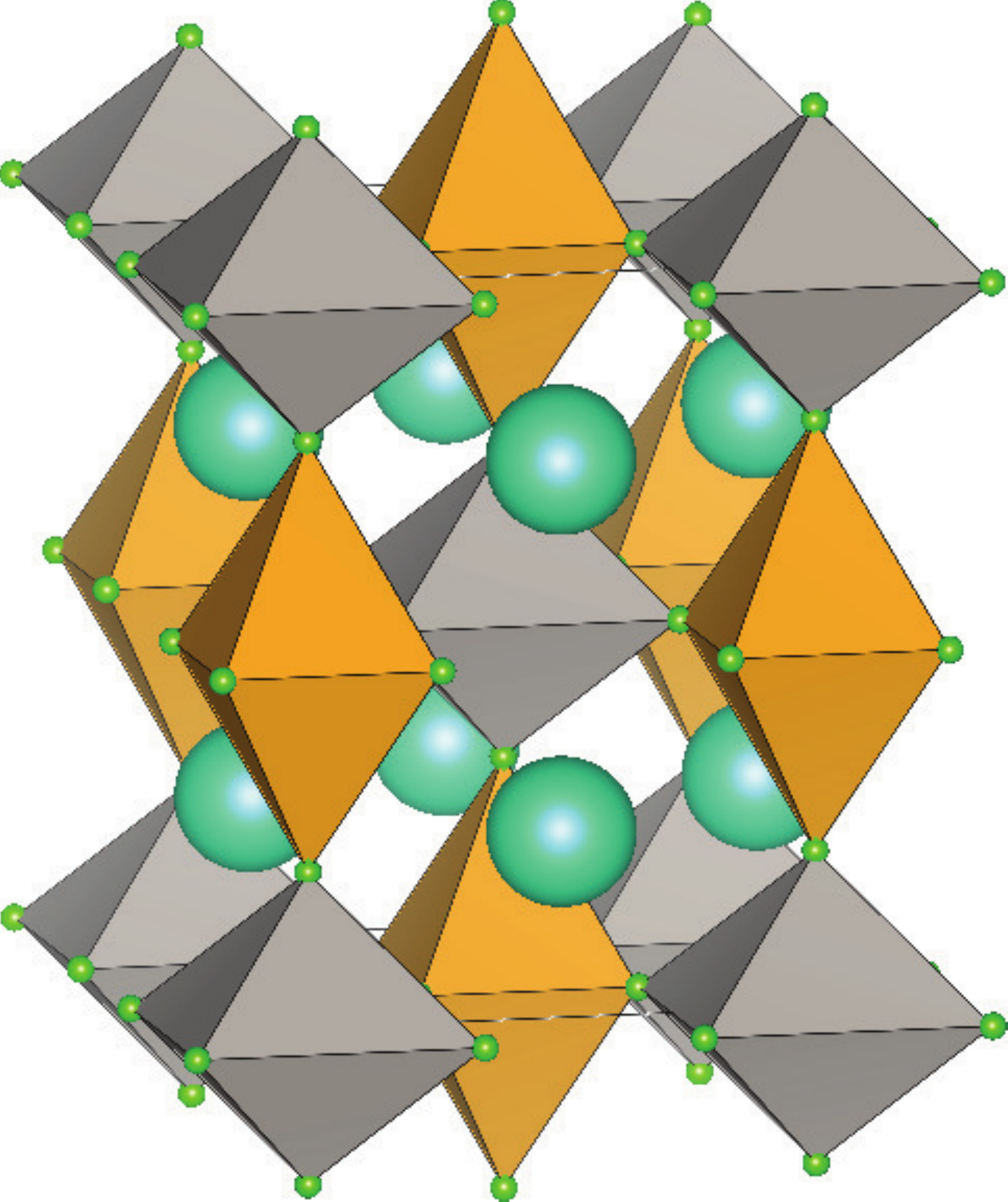}}
\caption{(Color online) Crystal structure of CsAuCl$_3$. Cs and Cl
  atoms are shown as large and small spheres, respectively. The Au
  atoms are located in the centres of the deformed octahedra. The
  octahedra that are elongated along $c$ correspond to Au$^{3+}$, the
  octahedra that are compressed along $c$ and extended in the basal
  plane correspond to Au$^{1+}$. This picture was generated using {\tt
    VESTA}.\cite{Momma/Izumi:2011}}
\label{fig:csaucl3-struc}
\end{figure}

CsAuCl$_3$ crystallizes in the perovskite structure with strongly
distorted Cl octahedra (see
Fig.~\ref{fig:csaucl3-struc}).\cite{Denner/Schulz/DAmour:1979} The
Au$^{3+}$ and Au$^{1+}$ cations are distributed in a three-dimensional
checkerboard pattern over the octahedrally-coordinated cation sites.
Thereby, the Cl octahedra surrounding the Au$^{3+}$ cations are
strongly elongated along the $c$ direction, whereas the Cl octahedra
surrounding the Au$^{1+}$ cations are compressed along $c$ and
extended within the basal plane.  This leads to an effective two-fold
linear coordination for Au$^{1+}$ and an effective four-fold
square-planar coordination for Au$^{3+}$.  The resulting structure has
tetragonal symmetry with space group $I4/mmm$.

Recent DFT calculations of CsAuCl$_3$ have found that the bands with
predominant Au$^{3+}$-$e_g$ character are lower in energy than the
corresponding $t_{2g}$-like
bands,\cite{Ushakov/Streltsov/Khomskii:2011} in contrast to the
``normal'' case of a TM ion in octahedral coordination, where the
$e_g$ orbitals are usually higher in energy than the $t_{2g}$ orbitals
(see e.g. the example of SrVO$_3$ discussed in
Sec.~\ref{sec_SrVO3}).  It has been argued that this reversal of the
crystal field splitting indicates a charge transfer character of
CsAuCl$_3$, with the 5$d$ states of the Au cations energetically lower
than the $p$ states of the Cl
ligands.\cite{Ushakov/Streltsov/Khomskii:2011} As a result, the
lower-lying \emph{bonding} energy bands, resulting from hybridization
between Au-$d$ and Cl-$p$ orbitals, have predominant Au-$d$ character,
in contrast to the more commonly found case where the TM-$d$-dominated
bands are \emph{antibonding} and are energetically higher than the
ligand $p$-dominated bonding states.  This change from antibonding to
bonding character of the nominal $d$ bands leads to a reversal of the
hybridization contribution to the $e_g$-$t_{2g}$ splitting, with the
$e_g$-dominated bands energetically lower than the $t_{2g}$-dominated
bands, since the $p$-$d$ hybridization is stronger for $e_g$ orbitals
than for $t_{2g}$ orbitals.  On the other hand, the purely
electro-static contribution to the $e_g$-$t_{2g}$ splitting is
independent of bonding or antibonding character and, within octahedral
coordination, should always lead to a higher energy of the $e_g$
orbitals compared to the $t_{2g}$ orbitals. The fact that the Au-$e_g$
bands are energetically lower than the Au-$t_{2g}$ bands thus also
indicates that in CsAuCl$_3$ the hybridization contribution of the
crystal field dominates over the electro-static contribution.

To validate this picture of the crystal field splitting in CaAuCl$_3$,
we now use the analysis described in the previous sections and
quantify the total crystal-field splitting of the Au$^{3+}$ cation as
well as the individual contributions corresponding to hybridization
and Coulomb interaction.

\begin{figure}
\includegraphics[width=\columnwidth]{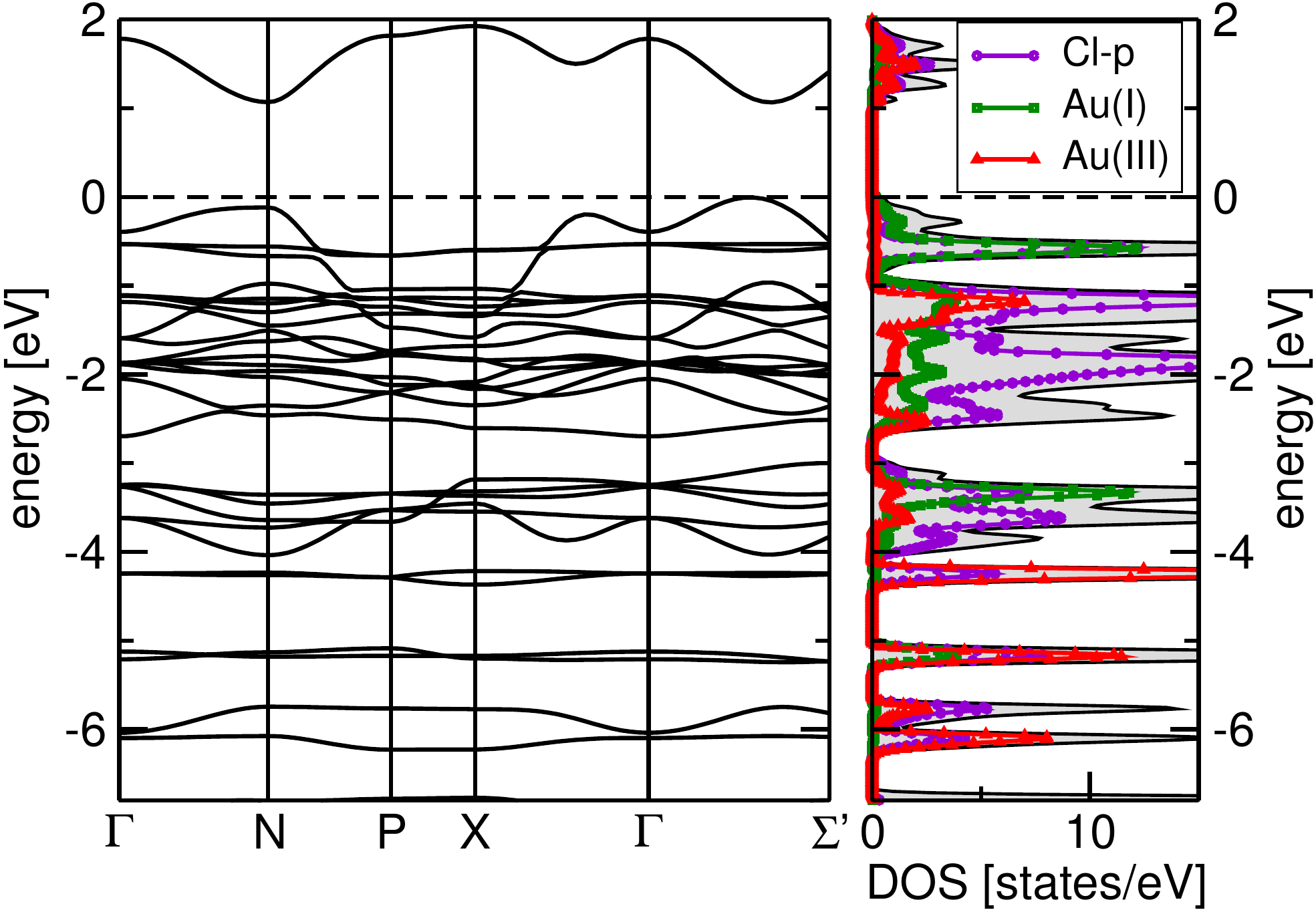}
\caption{(Color online) Left: calculated band structure of CsAuCl$_3$ along
  high-symmetry lines throughout the Brillouin zone. Right: orbital-
  and site-projected density of states (DOS) for CsAuCl$_3$. The total
  DOS is represented by the shaded region. The bottom of the energy
  gap is set to zero energy. $\Sigma'$ denotes the boundary of the
  first Brillouin zone along the $\Sigma$ line.}
\label{fig:CAC-dos}
\end{figure}

We calculate the electronic structure of CsAuCl$_3$ using the
experimental crystal structure reported in
Ref.~\onlinecite{Denner/Schulz/DAmour:1979} with $a=7.495$\,\AA,
$c=10.88$\,\AA, and Au$^{3+}$-Cl distances of 2.29\,\AA\ and
3.155\,\AA\ within the basal plane and along $c$. The resulting band
structure and density of states are shown in Fig.~\ref{fig:CAC-dos}.
The system exhibits an energy gap of approximately 1\,eV. The occupied
bands between $-$4.2\,eV and 0\,eV are dominated by the $p$ states of
the Cl anions and the $d$ states of the Au$^{1+}$ cation.  The $d$
states of the Au$^{3+}$ cation are mostly located at slightly lower
energies (between $-6.5$\,eV and $-4.2$\,eV), where they form rather
flat bands that also contain a significant amount of Cl-$p$ character
due to hybridization.

To obtain crystal-field energies, we construct three different sets of
MLWFs, which, in the notation of the previous sections, correspond to
``$d$''-, ``$dp$''-, and ``$dps$''-type Wannier functions:
\begin{enumerate}
\item[``$d$'':] Wannier functions constructed separately for each of
  the three groups of bands between $-6.5$\,eV and $-4.2$\,eV (each
  containing two bands). These six bands correspond to the five
  nominal Au$^{3+}$-$d$ bands, resulting from hybridization with the
  surrounding ligands, plus one band with strong Au$^{1+}$-$3z^2-r^2$
  character (at around $-5.2$\,eV).
\item[``$dp$'':] Wannier functions constructed simultaneously for all
  bands from $-6.5$\,eV up to 2\,eV, i.e. all nominal Cl-$p$ and
  Au-$d$ bands.
\item[``$dps$'':] Same as ``$dp$'', but in addition the Cl-$s$ bands
  (at around $-15$\,eV) are included.
\end{enumerate}

\begin{figure}
\includegraphics[width=\columnwidth]{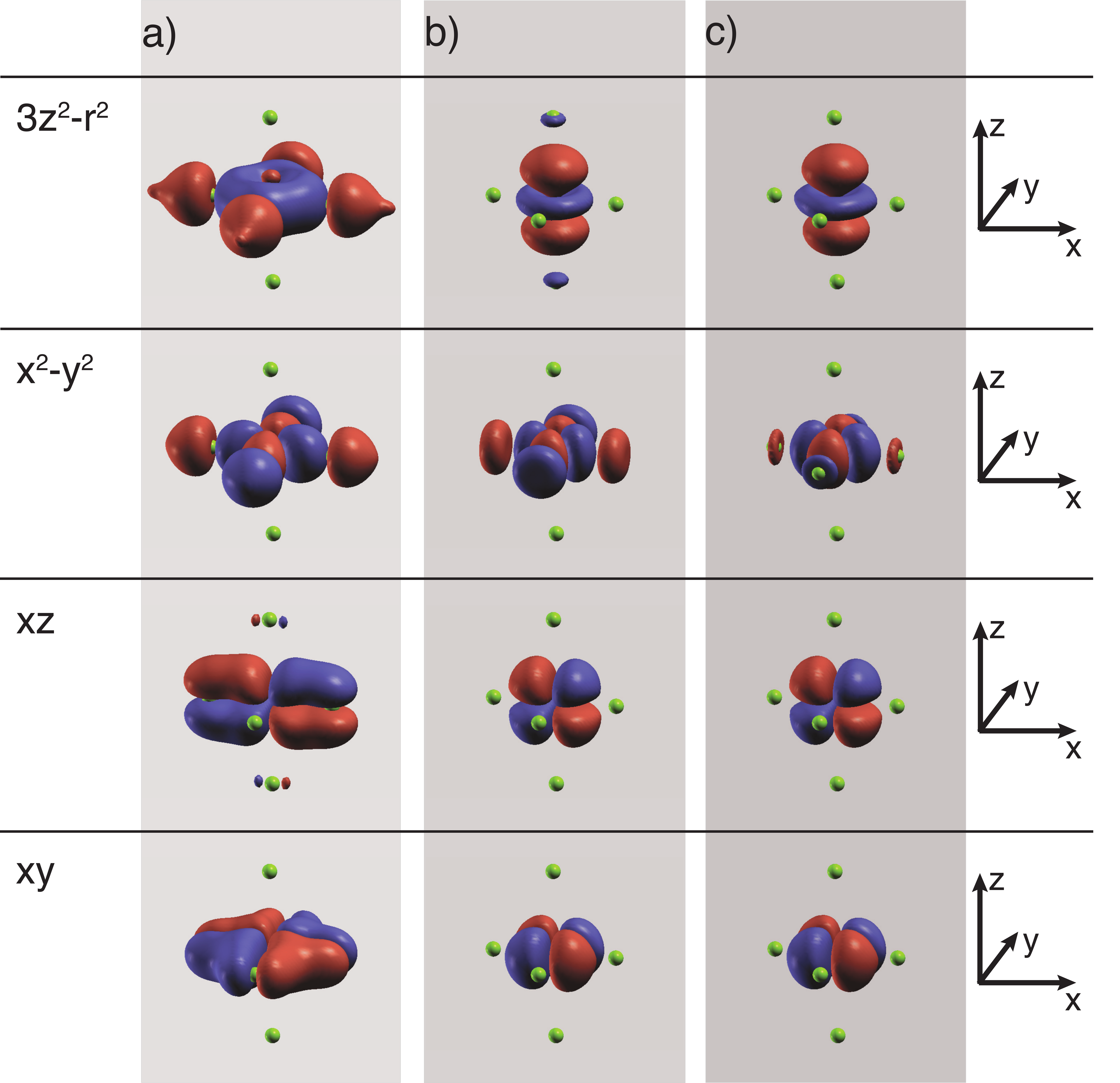}
\caption{(Color online) Symmetry inequivalent $d$-like MLWFs located
  on the Au$^{3+}$ site in CsAuCl$_3$, corresponding to the three
  different sets described in the text. Columns labeled (a), (b), (c)
  correspond to the $d$, $dp$ and $dps$ sets, respectively. The Cl
  atoms forming the octahedron around the Au$^{3+}$ site are indicated
  by small (green) spheres.}
\label{fig:CAC-mlwfs}
\end{figure}

Case ``$d$'' leads to five $d$-like Wannier functions centered at the
Au$^{3+}$ sites (plus one Au$^{1+}$-centered Wannier function) that
include hybridization with all the surrounding ligand states. These
orbitals are shown in Fig.~\ref{fig:CAC-mlwfs}(a). One can clearly
recognize that these Wannier functions correspond to \emph{bonding}
linear combinations of TM-$d$ and ligand-$p$ orbitals, since there are
no nodal planes in between the Au and the Cl positions. Due to the
long Au-Cl bond distance along the $c$ direction, the Au-$d$ orbitals
hybridize mostly with the four Cl ions within the basal plane. In case
of the ``$dp$'' set of MLWFs, shown in Fig.~\ref{fig:CAC-mlwfs}(b),
the hybridization with the Cl-$p$ ligand orbitals is removed from the
Au-$d$ Wannier functions, since the Cl-$p$ orbitals now appear as
separate Wannier functions within the set (only the Au-$d$-type
Wannier functions are shown in Fig.~\ref{fig:CAC-mlwfs}). Similarly,
in case ``$dps$'' (Fig.~\ref{fig:CAC-mlwfs}(c)) the hybridization
between Au-$d$ and Cl-$s$ is minimized in addition to the
Au-$d$-Cl-$p$ hybridization.

We note that due to the strong overlap between \mbox{Au$^{1+}$-$d$}
and Cl-$p$ bands, it is not possible to construct a set of
Au$^{1+}$-like Wannier functions analogous to the set ``$d$'' for the
Au$^{3+}$ Wannier functions.  Therefore, the different contributions
to the crystal-field splitting cannot be separated for the case of the
Au$^{1+}$ cation.

\begin{figure}
\centerline{\includegraphics[width=0.7\columnwidth]{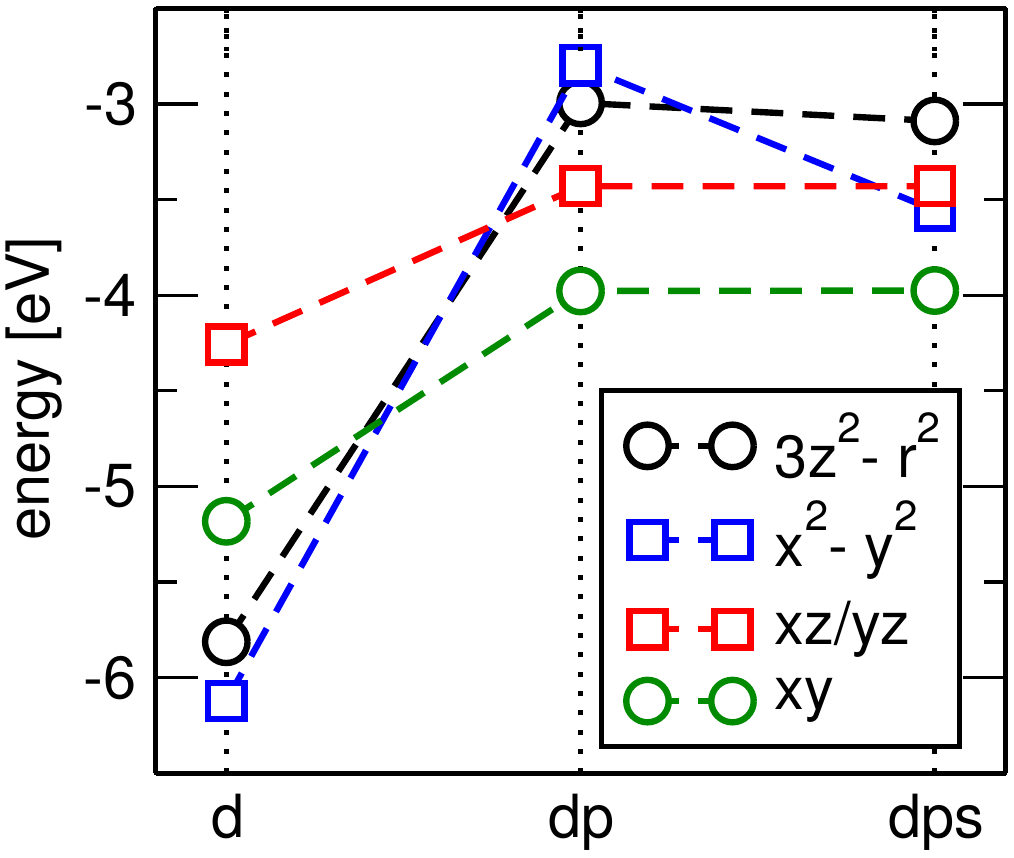}}
\caption{(Color on-line) On-site energies $\epsilon_m$ of the Wannier functions with
  predominant $d$ character on the Au$^{3+}$ site in CsAuCl$_3$ for
  the three different sets described in the text (``$d$'': fully
  hybridized, ``$dp$'': without $d$-$p$ hybridization, ``$dps$'':
  without $d$-$p$ and $d$-$s$ hybridization).}
\label{fig:CAC-e}
\end{figure}

Fig.~\ref{fig:CAC-e} shows the on-site energies of the
Au$^{3+}$-centered $d$-like Wannier functions for the three different
sets.  It can be seen that for set ``$d$'' (the most hybridized
Wannier functions) the $e_g$-like orbitals of the Au$^{3+}$ cations
are indeed lower in energy than the $t_{2g}$-like orbitals. These
energies essentially correspond to the centers of gravity of the
corresponding bands in Fig.~\ref{fig:CAC-dos}. We note that due to the
distortion of the octahedra, and the resulting tetragonal symmetry,
the two- and three-fold degeneracy among, respectively, the $e_g$ and
$t_{2g}$ orbitals is lifted.

Once the hybridization between the Au-$d$ and Cl-$p$ orbitals is
removed (or minimized), i.e. for the MLWFs of set ``$dp$'', the
energetic order between $e_g$ and $t_{2g}$ orbitals is reversed, and
the $e_g$-like Wannier functions now have higher energies than the
corresponding $t_{2g}$-like Wannier functions. Thus, it can be
seen that the effect of hybridization between Au-$d$ and Cl-$p$ states
is a \emph{downward} shift in energy of the Au-$d$ states, consistent
with a \emph{bonding} character of the nominal Au-$d$ bands as
described above.  This downward shift is much larger for the
$\sigma$-bonding $e_g$-like orbitals compared to the $\pi$-bonding
$t_{2g}$-like orbitals and thus the hybridization contribution to the
crystal-field places the $e_g$ orbitals lower in energy than the
$t_{2g}$ orbitals, i.e. the contribution of the $p$-$d$ hybridization
to the crystal-field splitting among the Au-$d$ orbitals is negative.

Comparing the orbital energies between sets ``$dp$'' and ``$dps$'',
one can see that hybridization with the Cl-$s$ states leads to an
\emph{upward} shift in energy of the $e_g$ orbitals, i.e. the
corresponding contribution to the ligand-field splitting is positive
since the Cl-$s$ states are energetically much lower than the Au-$d$
states.  One can also recognize that the hybridization with the Cl-$s$
states is weaker for the $3z^2-r^2$-like $d$ orbital compared to the
corresponding $x^2-y^2$-like orbital.  This is due to the quasi-planar
coordination of the Au$^{3+}$ cations resulting from the strong distortion of
the surrounding octahedra.

Our calculations thus confirm the above-described picture of the
electronic structure of CsAuCl$_3$ as a negative charge-transfer
system with inverted crystal-field splitting between $e_g$ and
$t_{2g}$ states of the Au$^{3+}$ cation. This reversed crystal field
splitting is due to the bonding character of the $d$-$p$ hybridization
in the nominal $d$ bands. Furthermore, our analysis reveals the
different character of the $d$-$p$ and $d$-$s$ hybridization, with the
much stronger $d$-$p$ hybridization dominating the resulting level
ordering. If one compares the average energy of the $e_g$-like
orbitals with the average energy of the $t_{2g}$-like orbitals of the
$dps$ set, i.e. with minimal hybridization, one can also see that the
``electro-static'' contribution to the crystal field, while being
rather small compared to the hybridization contributions, has the
expected sign and places the $e_g$ levels energetically higher than
the $t_{2g}$ levels. However, we note that, similar to the cases
discussed in the previous section, the tetragonal splitting between
the $3z^2-r^2$-like and $x^2-y^2$-like Wannier functions in set
``$dps$'' does not agree with the behavior expected from a simple
point charge model.  Within such a model one would expect the
$3z^2-r^2$ orbital to be lower in energy compared to the $x^2-y^2$
orbital for an octahedron that is elongated along $z$, since the
Coulomb repulsion with the ligands along $z$ is weakened. This
discrepancy can be explained by similar considerations as in
Sec.~\ref{sub_series}.

\section{Summary and Conclusions}

In summary, we have shown how MLWFs, constructed for different sets of
bands, can be used to separate different contributions to the crystal
field splitting of the TM $d$ electrons in TM oxides, provided that
the \emph{bonding} and \emph{antibonding} bands resulting from the
hybridization between the TM cation and the ligand anions are
energetically separated. The maximum localization condition then
allows separation of the orbital contributions located on different ions and
thus allows construction of Wannier functions corresponding to different
levels of hybridization.

We have demonstrated this approach using the example of cubic
perovskite SrVO$_3$ and two (hypothetical) series of tetragonally
distorted perovskite TM oxides. In all cases we could show that not
only the hybridization with the surrounding ligand $p$ orbitals, but
also that with the corresponding $s$ orbitals, gives sizable contributions
to the ligand field splitting. Furthermore, we have seen that in the
tetragonally distorted systems the remaining ``electro-static''
contribution to the splitting between the $3z^2-r^2$ and the $x^2-y^2$
orbital as well as between the $xz$/$yz$ and $xy$ orbitals changes
sign across the series and therefore does not agree with the
expectation based on a simple point charge model. We have discussed
several possible reasons for this discrepancy, perhaps the most
important being the difference between the pure ``crystal field
potential'' and the actual Kohn-Sham potential used to evaluate the
level splittings. We have also noted some conceptual difficulties in
the general definition of such a purely electro-static contribution to
the ligand field splitting, and we have used the level splittings
corresponding to the most localized Wannier functions with minimal
hybridization as a working definition for the electro-static level
splittings. 

Applying this approach to the charge-disproportionated 5$d$ system
CsAuCl$_3$ has allowed us to clearly separate the different
contributions to the level splittings on the Au$^{3+}$ site, thereby
demonstrating the competing tendencies between the $d$-$p$
hybridization on the one side and the $d$-$s$ hybridization and the
electro-static part on the other side. Our analysis thus confirms the
classification of Ref.~\onlinecite{Ushakov/Streltsov/Khomskii:2011} of
CsAuCl$_3$ as a negative charge transfer material with inverted
$e_g$-$t_{2g}$ splitting.

Finally, we note that the crystal-field splitting as discussed in this
work is by definition an orbital-dependent quantity, not a materials
constant. (In contrast, the true excitation energies of a material do
of course not depend on a specific orbital basis.)  Different choices
of Wannier functions (e.g. based on orbital projections such as those used in
Ref.~\onlinecite{Ku_et_al:2002} versus maximally localized) will lead
to somewhat different values for the corresponding splittings. Such
orbital dependence, however, is a rather common feature whenever one
attempts to interpret features of the electronic structure in a
chemical or TB-like picture.  Nevertheless, the observation of trends,
i.e. changes of the calculated splittings under some ``control
parameter'', can give valuable insights into the underlying mechanisms,
and lead to a better understanding of materials properties.

\begin{acknowledgments}
This work was supported by ETH Zurich and the Swiss National Science
Foundation.
\end{acknowledgments}

\appendix

\section{Analysis of the consistency of TB parameters obtained using MLWFs.}
\label{App_MLWFCons}

As mentioned in the introduction, the inherent non-uniqueness of
Wannier functions does not allow the extraction of TB parameters in a
unique way. Therefore, apart from the resemblance to atomic orbitals,
there is no apparent reason to use MLWFs instead of other possible
choices for the unitary matrices ${\rm U}^{(\vec{k})}$ in
Eq.~(\ref{eq:wf}). However, here we show that, at least for the
present case, our choice of MLWFs leads to consistency between the TB
parameters obtained for different sets of bands, and is perhaps even
slightly preferable to projection-based Wannier functions.

We first consider the simple case of a TB model for the perovskite
structure which includes only TM-$d_{xy}$ orbitals and the
corresponding $\pi$-oriented O-$p$ orbitals. Moreover, we restrict
ourselves to a model in which hoppings between sites beyond nearest
neighbors are negligible and consider the ``normal'' case where the
on-site energy of the TM-$d_{xy}$ orbital is higher than that of the
O-$p_\pi$ orbital.  The energy dispersion of the ``$d$-band'' within
such a TB model is:
\begin{equation}
\label{eq_disp1}
\epsilon_{\vec{k}}= \frac{(\varepsilon_{xy} + \varepsilon_{p})+
  \sqrt{(\varepsilon_{xy} - \varepsilon_{p})^2+ 8t^2 (2- \cos k_x -
    \cos k_y)} }{2} \quad .
\end{equation} 
Here, $\varepsilon_{xy}$ and $\varepsilon_{p}$ are, respectively, the
on-site energies of the $d_{xy}$ and $p_\pi$ orbitals, $t$ is the
hopping between these two orbitals, and $k_i$ is the component of the
wave vector along direction $i$.

Let us now consider another TB model which includes only effective
$d$-like Wannier functions, corresponding to antibonding orbitals
obtained by hybridization between the O-$p$ ligand states and the
TM-$d_{xy}$ orbitals. In this case the band dispersion is:
\begin{equation}
\label{eq_disp2}
\tilde{\epsilon}_{\vec{k}} = \tilde{\varepsilon}_{xy} + 2 \tilde{t} (
\cos k_x + \cos k_y) \quad ,
\end{equation}
where $\tilde{\varepsilon}_{xy}$ and $\tilde{t}$ are, respectively,
the effective on-site energy and nearest neighbor hopping amplitude of
the ``hybridized'' $d_{xy}$-type Wannier function (again we consider
only hopping between nearest neighbors).

In the limit where the difference between the on-site energy of the
$d_{xy}$-like TM state and the O-$p$-like ligand state is large
compared to the corresponding hopping amplitude,
i.e. $|t/(\epsilon_{xy} - \epsilon_{p})| \ll 1$, Eq.~(\ref{eq_disp1})
reduces to Eq.~(\ref{eq_disp2}) provided that
\begin{equation}
\label{eq_effModel}
\tilde{\varepsilon}_{xy} = \varepsilon_{xy} + \frac{ 4 t ^2
}{\varepsilon_{xy} - \varepsilon_{p}}, \ \ \ \tilde{t} = - \frac{ t^2
}{\varepsilon_{xy} - \varepsilon_{p}}.
\end{equation}

To test whether the Wannier functions obtained in
Secs.~\ref{sec_SrVO3} and \ref{sub_series} provide a consistent TB
description, we perform the following comparison. We evaluate the
right sides of Eqs.~(\ref{eq_effModel}) using the corresponding
on-site energies and hopping amplitudes of the set of Wannier
functions where both the nominal O-$p$-bands and TM-$d$-bands are
included in the construction of the Wannier functions,
i.e. $\varepsilon_{xy}=\varepsilon^{(dp)}_{xy}$,
$\varepsilon_{p}=\varepsilon^{(dp)}_{p_\pi}$, and
$t=t^{(pd)}_{xy,p_\pi}$. Then we compare the so-obtained
$\tilde{\varepsilon}$ and $\tilde{t}$ with the on-site energy
($\varepsilon^{(d)}_{xy}$) and nearest-neighbor hopping
($t^{(d)}_{xy,xy}$) obtained for $d_{xy}$-like Wannier functions when
only the effective $d$ bands are included in the Wannier construction.

\begin{figure}
\centerline{\includegraphics[width=0.95\columnwidth]{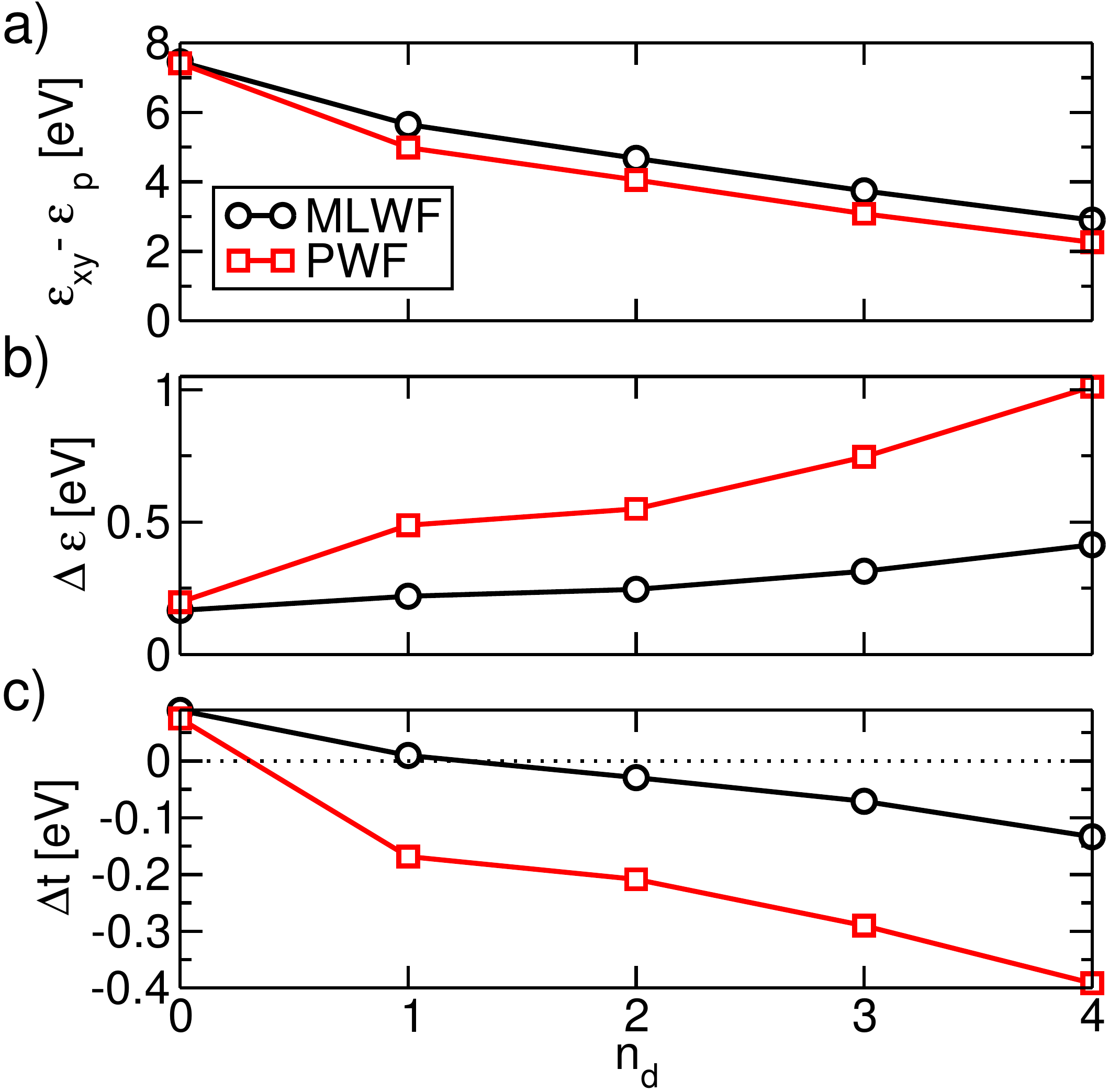}}
\caption{Values of $\varepsilon_{xy} - \varepsilon_p$ (panel a), $\Delta \varepsilon$ (panel b) and $\Delta t$
  (panel c), as defined in the text, obtained using MLWFs (black
  circles) and PWFs (red squares) as a function of $n_d$ for the series
  Tb$M$O$_3$ with $n_d \leq 4$.}
\label{fig:ConsPlot}
\end{figure}

Fig.~\ref{fig:ConsPlot} shows the differences \mbox{$\Delta
  \varepsilon = \varepsilon^{(d)}_{xy} - \tilde{\varepsilon}$} and
\mbox{$\Delta t = t^{(d)}_{xy,xy} - \tilde{t}$} (panel b and c,
respectively) for Tb$M$O$_3$ as function of the formal $d$ occupation
$n_d$. For comparison we also construct ``$d$''-type and ``$dp$''-type
sets of Wannier functions from projections on the relevant atomic
orbitals without subsequent spread minimization, i.e. we construct
so-called \emph{projector-based Wannier functions}
(PWFs).\cite{Ku_et_al:2002} The corresponding data are also included
in Fig.~\ref{fig:ConsPlot}. For both the MLWFs and the PWFs, $\Delta
\varepsilon$ decreases as the atomic number is decreased and the
energy separation between TM-$d$ and O-$p$-bands increases as shown in
Fig~\ref{fig:ConsPlot}a. A similar trend holds for $\Delta t$ with the
exception of $n_d=0$, i.e. TbScO$_3$. In TbScO$_3$ the $d$ bands
become strongly entangled with other bands at slightly higher
energies, which results in a certain ``discontinuity'' across the
series.

Thus, both types of Wannier functions provide a consistent TB
description in the sense that in the limit of large $d$-$p$ energy
separation the effective $d$-only TB model seems to become equivalent
to the more elaborate $d$-$p$ TB model. However, the discrepancy
between the two TB models is always slightly larger for the parameters
obtained from the PWFs compared to the case of the MLWFs.  Although
Fig.~\ref{fig:ConsPlot} seems to point to a better consistency of
MLWFs for calculating TB parameters, we stress that this
result might hold only for the considered class of compounds
(i.e. Tb$M$O$_3$) and a survey of additional chemistries and
structures would be desirable in future work.

\bibliography{references}

\end{document}